# Data-driven discovery of quasi-disordered mechanical metamaterials failed progressively

A. S. Bhuwal[a], Y. Pang[a], I. Ashcroft [b], W. Sun[a], T. Liu[a,*],

[a]Composite Research Group, Faculty of Engineering, University of Nottingham, University Park, Nottingham NG7 2RD, UK

[b]Centre for Additive Manufacturing, Faculty of Engineering, University of Nottingham, University Park, Nottingham NG7 2RD, UK

**Abstracts**

Natural cellular materials, such as honeycombs, woods, foams, trabecular bones, plant parenchyma, and sponges, may benefit from the disorderliness within their internal microstructures to achieve damage tolerant behaviours. Inspired by this, we have created quasi-disordered truss metamaterials (QTMs) via introducing spatial coordinate perturbations or strut thickness variations to the perfect, periodic truss lattices. Numerical studies have suggested that the QTMs can exhibit either ductile, damage tolerant behaviours or sudden, catastrophic failure mode, depending on the distribution of the introduced disorderliness. A data-driven approach has been developed, combining deep-learning and global optimization algorithms, to tune the distribution of the disorderliness to achieve the damage tolerant QTM designs. A case study on the QTMs created from a periodic Face Centred Cubic (FCC) lattice has demonstrated that the optimised QTMs can achieve up to 100% increase in ductility at the expense of less than 5% stiffness and less than 10% tensile strength. Our results suggest a novel design pathway for architected materials to improve damage tolerance.

Natural cellular materials, such as honeycombs, woods, foams, trabecular bones, plant parenchyma, and sponges, have inspired the development of mechanical metamaterials with desired or extreme mechanical properties [1–5]. These include various truss-like micro-lattices, i.e., truss mechanical metamaterials, at a scale ranging from nanometres to millimetres, manufactured using various additive manufacturing techniques [6–8]. Truss metamaterials have provided unique opportunities to create lightweight structural components of high performance, such as lightweight sandwich structures [9,10]. In addition, truss metamaterials are highly tailorable and can be designed to meet various multifunctional requirements, such as simultaneously loading bearing, active cooling and noise reduction [11,12].

Up till now, the majority of the relevant research focuses on the truss mechanical metamaterials of highly ordered structures, i.e., the bulk metamaterial is formed by repeating a representative volume element (RVE) in the two-dimensional (2D) or the three-dimensional (3D) space [13,14]. However, while nature provided cellular materials resemble truss lattice structures of ordered, periodic arrangement, they are not perfectly periodic. Spatial disorderliness has been observed in a wide range of natural cellular materials. Egmond, et al [15] has recently measured the disorderliness of the biological materials from trabecular bone to plant stems and fungi, using a disorder parameter $\dot{g}$ with $\dot{g} = 1$ representing the ordered system and $\dot{g} = 0.1$ the highly disordered system. They have identified the ranges of disorderliness within different types of biological materials, e.g. woods and fungi from $\dot{g} = 0.6$ to $0.8$; trabecular bone and dentin from $\dot{g} = 0.55$ to $0.65$; and corals and bee honeycomb from $\dot{g} = 0.9$ to $0.97$.

The role of disorderliness in mechanical performance for natural cellular materials has not been fully understood yet. Existing research has suggested that introducing disorderliness to periodic cellular materials can cause a reduction in stiffness, strength, ductility and fracture toughness [16–18]. However, Egmond, et al [15] have recently found that the disorderliness at the range of $\dot{g} = 0.6$ to $0.8$ within 2D Voronoi tessellation can cause an increase in toughness, through crack deflection, without loss of tensile strength in comparison with 2D regular hexagonal honeycombs. Based on this, they have hypothesized that structural disorder in natural cellular materials is a toughening mechanism and there may be a certain optimal degree of disorderliness in biogenic cellular materials in order to achieve damage tolerant behaviours. Here, we hypothesize that not only the level of disorderliness but also the distribution of disorderliness within natural cellular materials may play important role in achieving damage tolerance. As we have shown in SI Appendix section S1, truss metamaterials with identical level disorderliness can fail with either sudden, catastrophic brittle mode or progressive ductile mode during uniaxial tension tests, owing to different distribution of disorderliness.

Structural materials of high performance are expected to have suitable ductility to (i) fail in a progressive manner that can give prior warning to failure events, and (ii) have good load bearing capacity with the presence of flaws. It has been reported that highly ordered, periodic truss metamaterials often exhibit a sudden, catastrophic failure mode - when loaded beyond the yield point, localized bands of high strain emerge, causing catastrophic collapse [19–21]. To date, there are very limited studies on design methodology to achieve damage tolerant designs for mechanical metamaterials. Pham et al. [22,23] have used the hardening mechanisms found in crystalline materials to develop damage-tolerant designs, primarily under compression. They have found that the disorderliness introduced to periodic truss metamaterials, by mimicking the microscale structure of crystalline materials—such as grain boundaries, precipitates, and phases, can lead to the designs of progressive failure mode.

Motivated by the hypotheses on the role of disorderliness in natural cellular materials, we aim to develop a data-driven framework to optimize the distribution of disorderliness to achieve damage tolerance. Our approach has focused on quasi-disordered truss metamaterials (QTMs), which were formed by introducing small disorderliness to (parent) periodic truss metamaterials. As reported by Wang and Sigmund [24], the quasi-disordered mechanical metamaterials can be designed to achieve the extreme maximum isotropic elastic stiffness in the low-density limit and can preserve over 96% optimal stiffness at moderate densities up to 50%. We have employed deep-learning and optimization algorithms to tune the distribution of disorderliness within the parent periodic truss metamaterials to enable the resulted QTMs to preserve the original desired mechanical properties to the maximum extent.

Deep learning algorithms can be designed and trained to model highly nonlinear events with good accuracy and efficiency [25]. In solid mechanics, the Artificial Neural Networks (ANN) deep learning models have been employed to model complex material constitutive behaviours based on the data obtained by detailed finite element (FE) simulations or experimental measurements. Examples include modelling the nonlinear elastic behaviours based on the data created by FE simulations on representative volume elements [26,27]; predicting the stress states based on the applied strains in modeling path-dependent plasticity [28,29]; and predicting the evolution of the local strain distribution, plastic anisotropy, and failure during tensile deformation of a 3D-printed aluminium alloy using a feed-forward ANN framework [27]. The required amount of data for the deep learning process is problem dependent. For a complex material constitutive behaviour, millions of data may be required to achieve accurate modelling [29–31].

Our approach to seek the optimal QTMs follows 3-steps in the data-driven framework: 1) FE simulations to create a database of structural responses (outputs) corresponding to the database of geometric inputs; 2) The deep-learning ANN approach to establish the functional relationship that links inputs to outputs, and 3) non-gradient optimization to determine the optimal QTMs. Built upon the physical model and a custom-built loss function, numerical studies have suggested that our data-driven

**Significance**

Natural cellular materials are one of the most significant sources of inspiration for designing architected materials with unique behaviours. It has been recognised that the random disorderliness within the internal structures of natural cellular materials is a kind of toughing mechanism that fosters damage tolerance. However, little is known about how to incorporate random disorderliness into the designs of architected materials to achieve damage tolerance because of the infinite combinations of disorderliness. This work introduces the concept of quasi-disordered truss metamaterials (QTMs) by engineering-controlled degrees of disorderliness into the initially periodic material systems. Using a data-driven approach, disorderliness is tuned to enable QTMs to change failure mechanisms from brittle failure to progressive failure. We have demonstrated that higher ductility designs can be achieved with small expenses of stiffness and strength. This methodology can be generalized for designing any quasi-disordered metamaterial while minimizing ambiguity in indefinite solutions for highly complex structures.

approach may only require relatively small datasets.

**Creating the design space for quasi-disordered lattices**

Our approach to create QTMs of desired progressive failure modes was to introduce controlled (optimized) disorderliness to perfect periodic lattices of high performance. The periodic lattices with mechanical behaviour close to the Hashin-Shtrikman (H-S) theoretical limit [13,32,33], such as Face Centred Cubic (FCC), triangular, and Kagome lattices[18], were chosen to act as the parent periodic lattices. Built upon data-driven approaches, the distribution and level of the disorderliness were tuned through optimization procedures to ensure that the desired progressive failure modes could be achieved with maintaining or without much loss of the good mechanical properties inherited from the parent periodic lattices. The geometries of the QTMs in the design space were numerically created through two distinct approaches, i.e., (1) random perturbation of the spatial coordinates of the nodes of a parent periodic lattice; and (2) random strut thickness variation of a parent periodic structure. Consider a two-dimensional (2D) parent periodic lattice with $(x^i, y^i)$ representing the spatial coordinates of the $i$th node and $t^j$ the thickness of $j$th strut. To create a QTM through random perturbation of the spatial coordinates of the nodes, the perturbation $(\Delta x^i, \Delta y^i)$ was defined as [21]

$$\Delta x^i = \bar{x}^i - x^i = \alpha r \beta$$
$$\Delta y^i = \bar{y}^i - y^i = \alpha r \beta \tag{1.1}$$

Alternatively, to create a QTM through random strut thickness variation, the thickness of $j$th strut was defined as

$$\bar{t}^j = (1 + \gamma \beta) t^j \tag{1.2}$$

In Equations (1.1) and (1.2), $\beta$ $(-1 \leq \beta \leq +1)$ denotes a random variable following a uniform probability distribution; $r$ is the minimum distance between two nodes within the parent periodic lattice; $\alpha$ and $\gamma$ are the degrees of irregularity for the spatial perturbation and strut thickness variation, respectively. In this paper, small values are chosen for $\alpha$ and $\gamma$, e.g., $\alpha, \gamma = 0.1, 0.2, \text{ and } 0.3$, which leads to QTMs. The possible 2D design spaces for QTMs are illustrated in Fig. 1. The method introduced in the paper can be extended to three-dimensional (3D) QTMs, as shown in Fig.1.

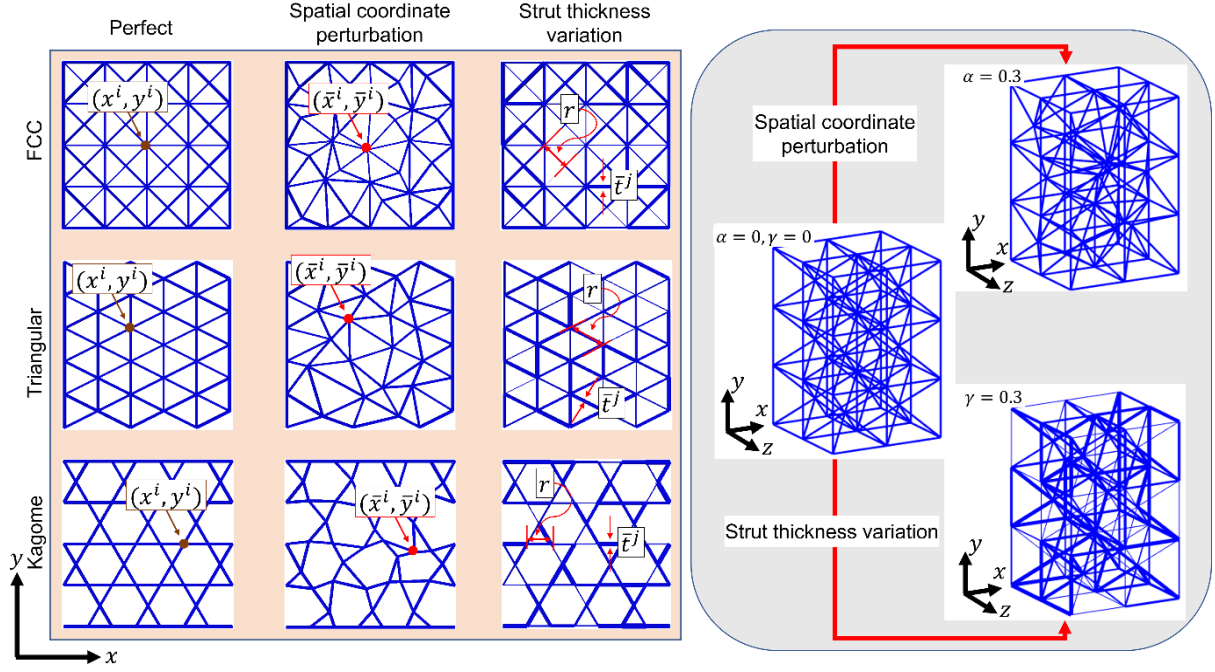

Figure 1. The QTM design space for two- and three-dimensional topologies

**A deep learning framework to map design space to output space**

The input and output databases were generated to feed into deep learning neural network for training purposes. The input database includes the geometric information of the QTM samples. Let $h$ denote the number of QTM samples included in the input database, and each QTM has $p$ nodes and $q$ struts. As shown in Fig. 2, for the $m$th QTM sample, $m = 1, 2,\ldots, h$, the geometric information included in the input database consists of (1) the perturbation of the spatial coordinates of the nodes, $\left(\Delta \mathbf{x}^m, \Delta \mathbf{y}^m\right)$, or (2) strut thickness variation, $\bar{\mathbf{t}}^m$, with

$$\Delta \mathbf{x}^m = \left[\Delta x^1,...,\Delta x^p\right]^{m\ \mathrm{T}},\quad \Delta \mathbf{y}^m = \left[\Delta y^1,...,\Delta y^p\right]^{m\ \mathrm{T}} \tag{1.3}$$

and

$$\bar{\mathbf{t}}^m = \left[\bar{t}^1,...,\bar{t}^q\right]^{m\ \mathrm{T}} \tag{1.4}$$

The output database includes the information on the structural responses of the QTM samples obtained by finite element (FE) simulations (details of FE modelling have been given in SI Appendix, section S2). As the current research focuses on the structural response under uniaxial tension, the normalized macroscopic stress data $\boldsymbol{\sigma}^m = \left[\sigma^1, \sigma^2, ..., \sigma^n\right]^{m\ \mathrm{T}}$ collected at a sequence of $n$ predefined, equally spaced normalized macroscopic uniaxial strains, $\boldsymbol{\varepsilon}^m = \left[\varepsilon^1, \varepsilon^2, ..., \varepsilon^n\right]^{m\ \mathrm{T}}$, were stored in the output database for the $m$th QTM sample. Here, the macroscopic stresses are defined as the stresses averaged over the entire model, and the macroscopic tensile strain is defined as the elongation over the original length of the model.

A feed-forward deep-learning ANN can be trained, using the input and output databases, to map the functional relationship between the input and output databases, as shown in Fig.2, i.e.,

$$\boldsymbol{\sigma} = \left[\sigma^1, \sigma^2, ..., \sigma^n\right]^{\mathrm{T}} = f_1(\Delta \mathbf{x}, \Delta \mathbf{y}), \text{ or}$$
$$\boldsymbol{\sigma} = \left[\sigma^1, \sigma^2, ..., \sigma^n\right]^{\mathrm{T}} = f_2(\overline{\mathbf{t}}) \tag{1.5}$$

Neural network architecture refers to assembling neurons into layers: Each neuron uses a mathematical transformation of weights and biases to generate an output layer. For example, the mathematical formula of a feed-forward propagation neural network of $l$ layers can be written as:

$$\begin{aligned}
\mathbf{a}_1 &= g\left(\boldsymbol{\theta}^{[1]}\boldsymbol{\psi} + \mathbf{b}_1\right) \\
\mathbf{a}_2 &= g\left(\boldsymbol{\theta}^{[2]}\mathbf{a}_1 + \mathbf{b}_2\right) \\
&\ldots \\
\mathbf{a}_{ii} &= g\left(\boldsymbol{\theta}^{[ii]}\mathbf{a}_{ii-1} + \mathbf{b}_{ii}\right) \\
&\ldots \\
\mathbf{a}_{l-1} &= g\left(\boldsymbol{\theta}^{[l-1]}\mathbf{a}_{l-2} + \mathbf{b}_{l-1}\right) \\
\boldsymbol{\sigma} &= \boldsymbol{\theta}^{[l]}\mathbf{a}_{l-1} + \mathbf{b}_l
\end{aligned} \tag{1.6}$$

where $\boldsymbol{\theta}^{[ii]}$ is a weight matrix in the $ii$th layer, $ii = 1, 2, ..., l$; $\mathbf{b}_{ii}$ the bias vector in the $ii$th layer; $\mathbf{a}_{ii}$ the output vector in the $ii$th layer; $g$ the activation function; $\boldsymbol{\psi}$ is the input vector of the ANN, i.e.,

$$\boldsymbol{\psi} = \left[\left(\Delta x^1, \Delta y^1\right), ..., \left(\Delta x^p, \Delta y^p\right)\right]^{\mathrm{T}}, \text{ or}$$
$$\boldsymbol{\psi} = \overline{\mathbf{t}} = \left[\overline{t}^1, ..., \overline{t}^q\right]$$

The learning (training) procedure tunes the ANN components to minimise the cost function $J(\boldsymbol{\theta}^{[ii]}, \mathbf{b}_{ii})$, which is related to the loss function $\mathcal{L}(\boldsymbol{\sigma}_{pred}^m, \boldsymbol{\sigma}_{true}^m)$, by the following Equation [29]

$$J(\boldsymbol{\theta}^{[ii]}, \mathbf{b}_{ii}) = \frac{1}{h}\sum_{m=1}^{h}\mathcal{L}(\boldsymbol{\sigma}_{pred}^m, \boldsymbol{\sigma}_{true}^m) \tag{1.7}$$

where the loss function measures the accuracy of the trained ANN by evaluating the difference between the predicted stresses, $\boldsymbol{\sigma}_{pred}^m = \left[\sigma^1, \sigma^2, ..., \sigma^n\right]_{pred}^{m\,\mathrm{T}}$ and the real stresses, $\boldsymbol{\sigma}_{true}^m = \left[\sigma^1, \sigma^2, ..., \sigma^n\right]_{true}^{m\,\mathrm{T}}$. To improve the learning efficiency of the ANN model, the normalized macroscopic stress data $\boldsymbol{\sigma}^m = \left[\sigma^1, \sigma^2, ..., \sigma^n\right]^{m\,\mathrm{T}}$ for a QTM sample can be divided into three groups, which correspond to the three zones in the stress-strain relation for the QTM sample under uniaxial tension, respectively, as shown in Fig. 2. It is noted that the structure experiences (1) elastic deformation in Zone I, (2) plastic deformation caused by the failure of a limited number of struts in Zone II, and (3) final catastrophic failure in Zone III. Numerical experiments on quasi-disordered FCC lattices have suggested that the stress data in the three groups (Zones) have significantly different variances across the QTM samples (see SI Appendix, section S3). Based on this finding, a novel quantile regression loss function has been employed in this work, which is given as:

$$\mathcal{L}(\boldsymbol{\sigma}_{pred}^m, \boldsymbol{\sigma}_{true}^m) = \frac{1}{3n} \sum_{i=1}^{3} \left[ \sum_{\substack{k=1 \\ \sigma_{true}^k < \sigma_{pred}^k}}^{n} (\lambda_i - 1)\left(\sigma_{pred}^k - \sigma_{true}^k\right)^2 + \sum_{\substack{k=1 \\ \sigma_{true}^k \geq \sigma_{pred}^k}}^{n} \lambda_i \left(\sigma_{pred}^k - \sigma_{true}^k\right)^2 \right] \quad (1.8)$$

where $\lambda_i$, $i = 1, ..., 3$, are the chosen quantiles for the three groups of the stress data and have values between 0 and 1. The quantile loss function is an extension of the Mean Square Error (MSE) that has the quantile $\lambda_i = 0.5$. The larger the value $\lambda_i$, the more under-predictions are penalized than over-predictions. Our numerical experiments have suggested that it can help to improve deep-learning efficiency (see SI Appendix, section S3) and reduce the amount of data required for the deep-learning process by using distinct $\lambda_i$ values at different Zones.

**Non-gradient-based design optimization**

Design optimization procedures can be employed to tune the distribution of disorderliness within the parent periodic lattices to achieve desired progressive failure modes. Based on the trained deep-learning ANN models, the mathematical model for design optimization can be described as follows

**Objective function**:
Maximize:
$$T(\Delta \mathbf{x}, \Delta \mathbf{y}) \text{ OR } T(\bar{\mathbf{t}}) \quad (1.9)$$

**Constraints**:

$$\Delta x_{min} \leq \Delta x^i \leq \Delta x_{max};$$
$$\Delta y_{min} \leq \Delta y^i \leq \Delta y_{max}, \quad i \in [1, p]$$
or
$$t_{min} \leq \bar{t}^j \leq t_{max}, \quad j \in [1, q] \quad (1.10)$$
and
$$Max\left[\sigma^1, \sigma^2, ..., \sigma^n\right] \geq \sigma_{min}$$

where $\Delta x_{min}$, $\Delta y_{min}$, $t_{min}$ and $\sigma_{min}$ are the lower bounds of design variables; and $\Delta x_{max}$, $\Delta y_{max}$ and $t_{max}$ the upper bounds. In Equation (1.9), The objective function $T$ is a measurement related to the deformation capacity of the QTMs, such as ductility and strain energy density. The optimization problem described in Equation (1.9) can be solved using non-gradient based optimization algorithms such as the Genetic Algorithms [34,35], the Particle Swarm Optimization [36], and the Simulated Annealing (SA) Optimization [37].

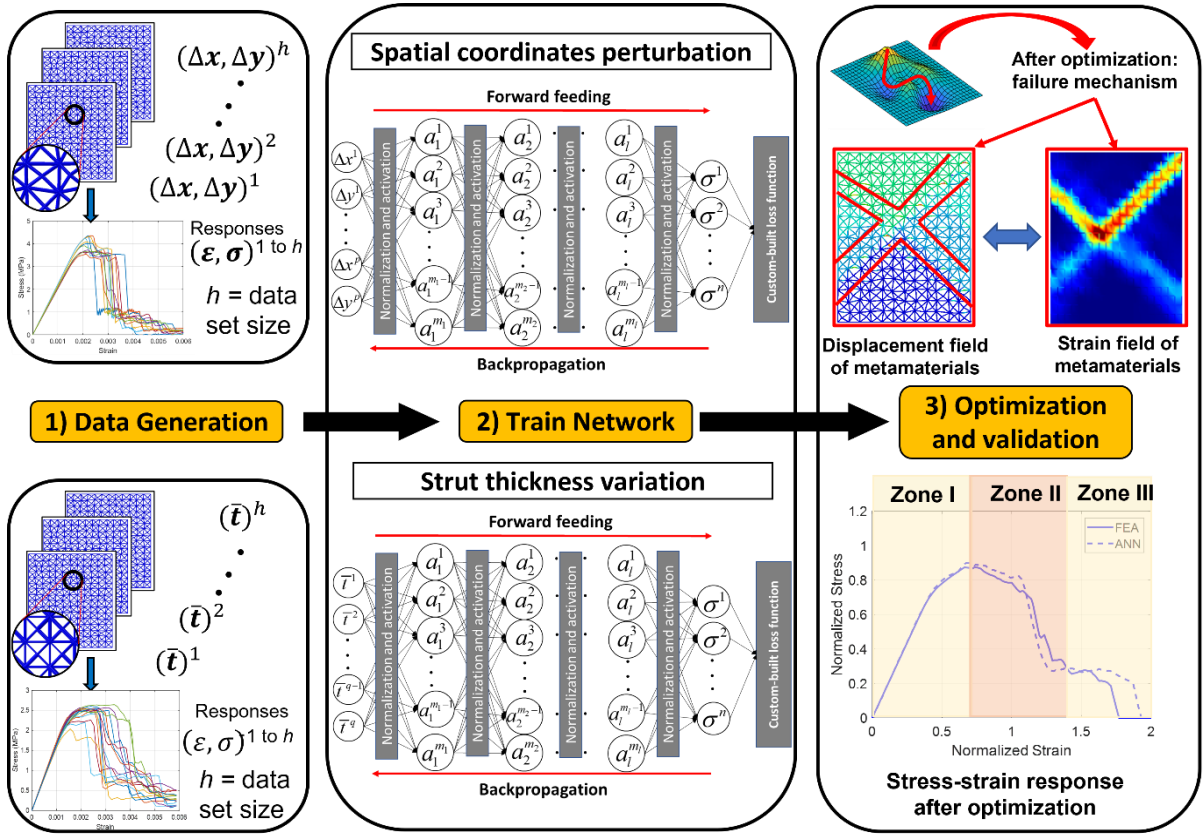

Figure 2 Overview of the methodology showing steps involved in designing QTMs, starting from step 1) data generation of spatial coordinate perturbation and strut size variation, step 2) ANN training with customised loss function to accurately map complex input and output variable, and step 3) optimization and validation of the designs.

**Results**

We demonstrate the success of the proposed method through the discovery of the high-performance 2D QTMs of the progressive failure modes. The QTMs were created based on a parent periodic FCC lattice (SI Appendix, section S1). It was assumed that the lattice was made of the aluminium alloy Al-1050A, and with relative density, $\bar{\rho} = 0.2$. This relative density value was chosen as the FE simulations had suggested that, at nodal perturbation $\alpha = 0.2$, the QTMs had a high variance in ductility and low variances in stiffness and strength (SI Appendix, section S4). The parent FCC lattice consisted of 12- and 16-unit cells periodically arranged along the $x$ and $y$ directions, respectively, with dimensions of 120 mm in $x$ direction and 160 mm in $y$ directions, see Fig. 3A. This geometry was chosen to ensure that the mechanical properties, i.e., macroscopic stiffness and peak strength were not sensitive to the size of the test samples (SI Appendix, section S5). The FE simulations have suggested that the parent lattice exhibits a sudden, catastrophic structural failure mode under uniaxial tension along the $y$ direction, as shown in Fig. 3B for the corresponding normalized macroscopic stress-strain relation. As shown in the insert of Fig. 3B for the distribution of the displacement at failure, the failure event was mainly caused by the formulation of a single shear band across the sample. For the macroscopic stress-strain relation shown in Fig. 3B and the rest of the paper, the stress values have been normalized by the peak stress, and the strain values by the maximum strain of the parent FCC periodic lattice.

**The ANN models**

The first ANN model was created based on the scenario in which disorderliness was introduced into the FCC periodic lattice via the perturbation of the spatial coordinates of the nodes. The geometries of 5000 QTM samples were generated with irregularity, $\alpha = 0.2$, at constant relative density $\bar{\rho} = 0.2$. The input database containing perturbation of the spatial coordinates of the nodes, $\boldsymbol{\psi} = \left[\left(\Delta x^1, \Delta y^1\right), ..., \left(\Delta x^p, \Delta y^p\right)\right]^T$, and the output database containing the normalized macroscopic stress data $\boldsymbol{\sigma} = \left[\sigma^1, \sigma^2, ..., \sigma^n\right]^T$ were created to train the ANN model. The ANN model consisted of 7 hidden layers with 4096, 2048, 1024, 1024, 1024, 512, and 512 neurons, respectively, in sequence from input to output layers. The numerical experiments have suggested that the structure of the ANN has achieved high efficiency in deep learning. The tuning of ANN model hyperparameters was obtained by performing Bayesian Optimization (SI Appendix, section S3), based on the loss function with $\lambda_1 = 0.5$, $\lambda_2 = 0.45$ and $\lambda_3 = 0.1$, respectively. The second ANN model was created based on the scenario in which the disorderliness was introduced into the parent FCC periodic lattice via strut thickness variation. The ANN model was trained based on the input database containing struct thicknesses, $\boldsymbol{\psi} = \bar{\mathbf{t}} = \left[\bar{t}^1, ..., \bar{t}^q\right]$, and the output database resulted from the FE simulations for 5000 QTM samples, with irregularity $\gamma = 0.1$ and at constant relative density, $\bar{\rho} = 0.2$. The ANN parameters are the same as in the previous case, except that the chosen quantiles were $\lambda_1 = 0.5$, $\lambda_2 = 0.5$ and $\lambda_3 = 0.3$, respectively. We trained the two ANN models for 100 epochs with an early stopping function when no improvements were made for ten iterations consecutively (the evaluations on the full dataset are presented in SI Appendix, section S3).

**The design optimization**

Ductility and strain energy density were employed as the objective functions of the optimization model. Here and throughout the rest of the paper, the ductility is defined as the macroscopic tensile strain at failure, which corresponds to the post-peak macroscopic stress equivalent to 25% of the peak macroscopic tensile stress; and the strain energy density was calculated as the area under the macroscopic stress-strain curve.

The objective functions were optimized with the constraints of allowable nodal perturbation $\alpha = 0.2$, allowable struct thickness variation $\gamma = 0.1$, and minimum normalized strength $\sigma_{\min} = 0.9$ using the simulated annealing (SA) optimization algorithm (MATLAB [38]). The strength constraint can ensure that the resulted QTMs preserve more than 90% of the strength from the parent periodical FCC lattice.

The relative density of optimized QTMs have been found to be maintained at a constant value $\bar{\rho} = 0.2$. The optimization results are presented below for the QTMs having improved ductility owing to the progressive failure process.

**The optimization results**

Based on the first ANN model, we optimized the distribution of the perturbation of the spatial coordinates of the nodes for the maximised ductility design and the maximised strain energy density design, respectively, as shown in Fig. 3 for the optimized designs. The optimized distributions of the nodal perturbation were used to create the corresponding FE models for validation and interpretation purposes. Compared to the periodic FCC lattice (Fig. 3B), which failed in a sudden, catastrophic manner, the optimized designs exhibited progressive failure modes. The optimized design based on the maximised ductility design model exhibits a 73% increase in ductility (Fig. 3C); and the optimized design based on the maximised strain energy density model exhibits a 56% increase in strain energy density (Fig. 3G), both with less than 5% reduction in stiffness and up to 10% reduction in strength. The ANN predictions have a good agreement with the FE simulation results. FE simulations suggested that the progressive failure modes in the optimized designs were mainly achieved by shear band branching that causes load-path shift to undamaged struts, as shown in Fig. 3F (QTM -N1) and Fig. 3J (QTM -N4). However, it has been found that the optimized design was not unique: the solution is sensitive to the initial distribution of the perturbation. This indicates that the method can generate different designs with similar local optima. To illustrate this, based on three different initial distributions that were randomly picked from the input dataset, we obtained the optimized distributions of the perturbation for the maximised ductility designs (Figs. 3 C, D and E for QTMs -N1, -N2 and -N3) and the maximised strain energy density designs (Figs. 3 G, H and I for QTMs -N4, -N5 and -N6), respectively. Albeit slight differences in mechanical behaviours, these optimized designs all show progressive failure modes with a significant increase in ductility or strain energy density compared to the periodic FCC lattice. The inserted distribution of the displacement at failure, as shown in Figs. 3 D, E, F and Figs. 3 H, I, J, have suggested that the progressive failure modes were mainly caused by shear band branching in different patterns.

Next, based on the second ANN model, we optimized the variation of strut thickness within the parent periodic FCC lattices using the ductility objective function, as shown in Fig. 4 for the three optimized designs. As in the previous case, we obtained the optimum designs that exhibited progressive failure modes compared to the periodic FCC lattice (Fig. 3A). The optimized designs exhibit more than 80% increase in ductility with the expense of less than 10% strength and less than 5% stiffness (Figs. 4A, B and C); and again the progressive failure modes in the optimized designs were mainly achieved by the shear band branching with different patterns, as shown in Fig. 4B, C and D of QTMs -S1, -S2 and -S3.

The optimized designs based on strain energy density objective function exhibit similar behaviours, i.e., more than 61% increase in strain energy with the expense of less than 10% strength and less than 5% stiffness (see SI Appendix, section S6). Our results show that the QTMs resulting from the strut thickness variation are more prone to brittle failure compared to those resulting from the spatial nodal perturbation.

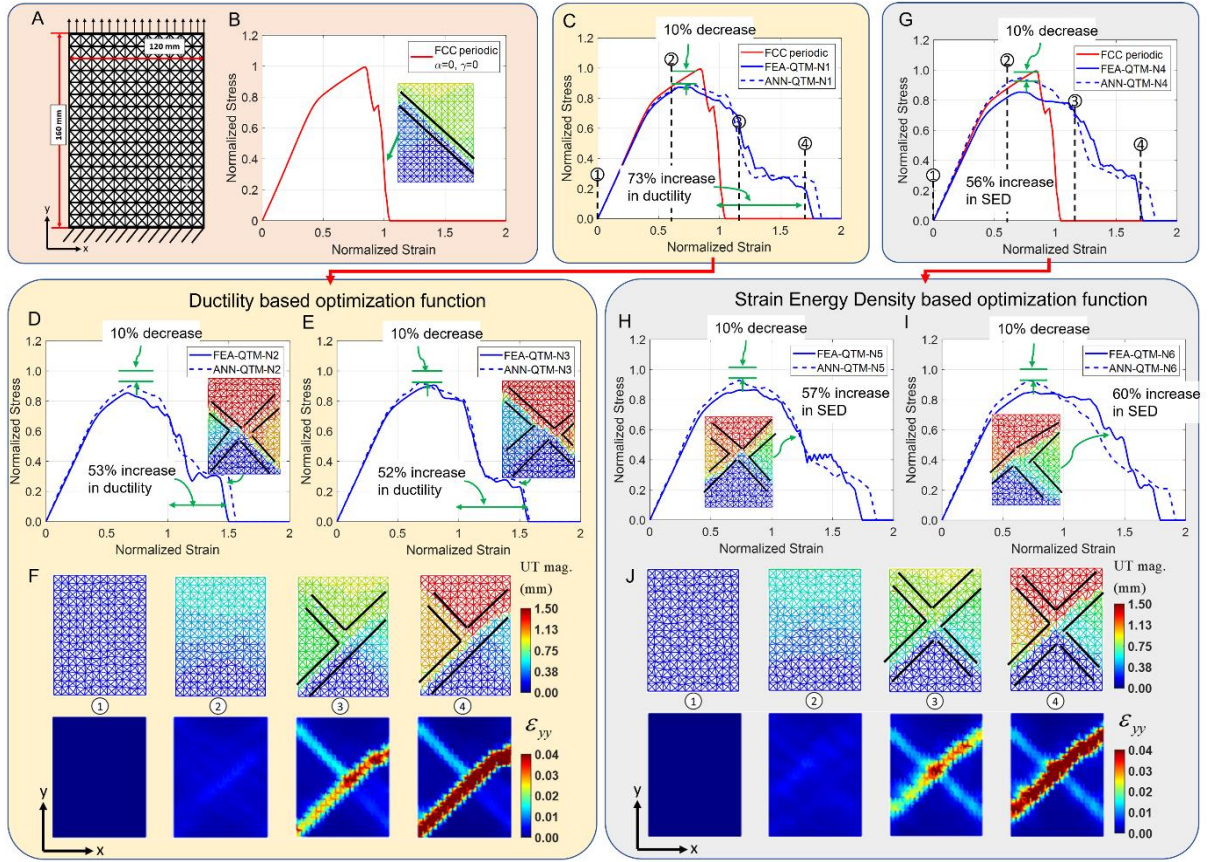

Figure 3 Designs of metamaterials based on spatial coordinate perturbations (A) the dimensions and boundary conditions of the FE model; (B) the normalized macroscopic stress-strain relation of a parent periodic FCC lattice; (C, D and E) the normalized stress-strain curves of three optimized QTMs (-N1, -N2 and -N3, using ductility objective function) obtained by the FE simulations and ANN predictions; (F) the detailed distributions of displacements in the lattices along with the continuum plots of microscopic strain $\varepsilon_{yy}$ [21] at selected macroscopic strains of a QTM (-N1), showing shear band branching; the corresponding results of the three QTMs (-N4, -N5 and -N6) obtained using strain energy density objective function are shown in (G, H, I and J). The inserts of (D, E, H and I) show the distribution of the displacement at failure, which is caused by the formulation of the shear band branching with different patterns.

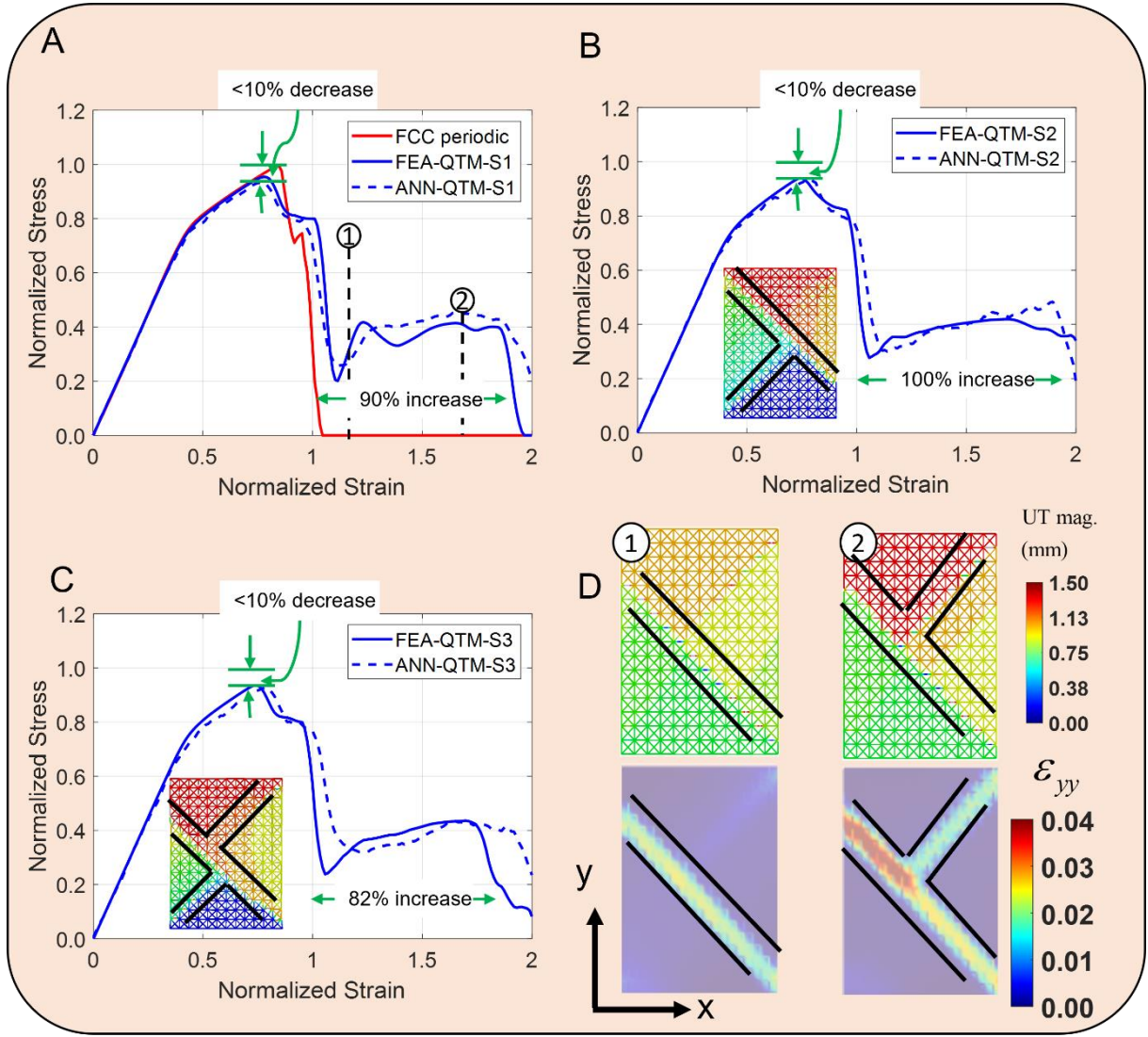

Figure 4 Designs of metamaterials based on strut size variation using ductility objective function, (A, B and C) the normalized stress-strain curves of three optimized QTMs (-S1, -S2 and -S3) obtained by the FE simulations and ANN predictions; (D) the distribution of microscopic strain $\varepsilon_{yy}$ at selected macroscopic strains of a QTM (-S1), showing that shear band branching causes progressive failure mode. The inserts of (B, and C) show the distribution of the displacement at failure, which is caused by the formulation of the shear band branching with different patterns.

**Experimental study**

To validate the methodology described above, test specimens were manufactured, using the L-PBF (Laser powder bed fusion) manufacturing technique, for uniaxial tension test based on three selected FCC lattice designs, i.e., the parent periodic FCC lattice, the QTM with progressive failure mode (QTM-1), and the QTM with sudden failure mode (QTM-2). The optimized design shown in Fig. 3J was used as the geometry of the QTM-1; and the geometry of the QTM-2 was selected from the data used for deep learning, which exhibited sudden catastrophic failure mode according to the FE simulations. To question if the failure modes of the lattices depend on the parent material, we chose polymers as the parent material instead of Aluminium alloy (Al1050A), which was used in the design optimization as described previously. Two types of polymers (i.e., *Polymer-1* and *Polymer-2)* were used as the parent materials, which were created by combining commercialized acrylic (Objet Vero-Clear FullCure810) and rubber-like material (Objet Tango-Gray FullCure950). *Polymer-1* has a mixture of 75% acrylic and 25% rubber-like material, and P*olymer-2* has a mixture of 50% acrylic and 50% rubber-like material. Both polymers show elastic-plastic stress-strain response, with *Polimer-2* being much more ductile than *Polymer-1* (see SI Appendix, section S7). The tensile tests were conducted at room temperature using a 0.1% strain rate using an INSTRON testing system, as shown in Fig .*5A* for a photograph of the experimental setup. Each specimen contained 12 x 16 cells, and the geometry of a specimen had the size of 120 mm x 160 mm (with the height of 25mm clamping at both top and bottom sides) with 1 mm out-of-plane thickness, and each strut had 0.4 mm thickness (the detailed geometry of the experimental sample is presented in SI Appendix, section S7). Digital image correlation (DIC) was employed to capture the full-field strain evolution of the samples during the full fracture process. A CCD camera (Thorlabs DCC1545M) with an imaging lens (100mm focal length) was configured at a spatial resolution of 5 pixels / mm and a frame rate of 20 fps. In the DIC algorism, the subset image was 128 x 128 pixels, and the step size was 64 pixels to maintain a high level of speckle correlation [39,40].

The normalized macroscopic stress-strain curves for the specimens made of *Polymer-1* and *Polymer-2* are shown in Figs. 5*B* and *F,* respectively. For both parent materials, the periodic FCC lattice failed in a sudden, catastrophic manner. Compared to the periodic FCC lattice, the QTM-1 achieved a 60% increase in ductility for *Polymer-1* and a 33% increase for *Polymer-2*, respectively, without significantly decreasing the mechanical stiffness ($\leq 3\%$) and the strength ($\leq 13\%$). On the other hand, for both parent materials, the QTM-2 exhibited sudden, catastrophic failure mode with ductility either slightly higher (Polymer-*1*) or much lower (*Polymer-2)* than that of the periodic FCC lattice.

To further examine the failure mechanisms, a series of video snapshots at four selected normalized macroscopic tensile strains, i.e. 1.5%, 2.8%, 3.0% and 4.5%, respectively, were presented in Fig. 5 C-E (*Polymer-1*) and Figs. 5G-I (*Polymer-2*). For both parent polymers, both the periodic FCC structure and QTM-2 fail instantaneously owing to a single shear band formation across the sample at the 3.0% normalized tensile strain. Interestingly, the shear band deflection in Fig. 5C did not lead to progressive failure mode owing to the brittle parent material (Polymer 1). On the other hand, the QTM-1 design develops damage-tolerant behaviours via progressive failure modes owing to shear band branching (Fig. 5*E*, Polymer-1) or excessive tortuosity in the development of the shear band (Fig. 5*I*, Polymer-2).

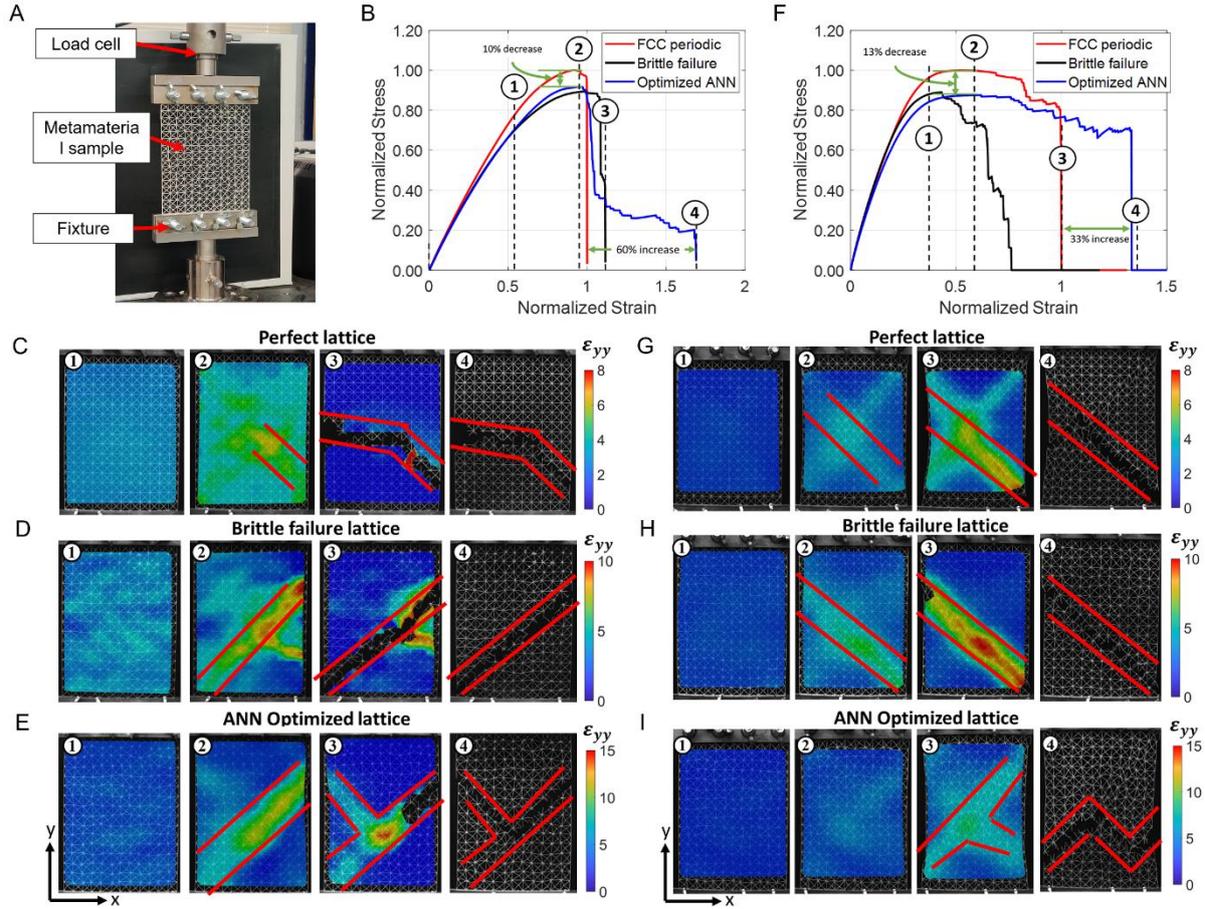

Figure 5 Tensile tests of the parent periodic FCC lattice, the QTM with progressive failure mode (QTM-1), and the QTM with sudden failure mode (QTM-2). (A) 3D printed QTM-1 using *polymer1* sample and experimental setup, (B) the normalized stress-strain curves of QTM-1 showing ~60% increase in ductility using *polymer1*, (C), (D) and (E) snapshots of *polymer1* samples at different global strains during the test of parent periodic FCC, QTM-2, and QTM-1, respectively. The DIC images show the progression of microscopic strain $\varepsilon_{yy}$ and shear band branching (F) the normalized stress-strain curves of QTM-1 showing ~33% increase in ductility using *polymer2*, (G), (H) and (I) are snapshots at different global strains for *polymer2* (the fourth snapshots of parent periodic FCC lattice, QTM-1 and QTM-2 were taken after the final fracture occurred)

## Discussions

The structures of the natural cellular materials exhibit a certain level of disorderliness. Prior to this work, it was well established that the disorderliness within cellular materials can cause a reduction in stiffness, strength, ductility, and fracture toughness. This has been demonstrated by a range of theoretical and experimental studies by Romijn et al. [16], Chen et al. [17], Tankasala et al. [18], and Xu et al. [41]. However, in this paper, we have shown that the level and the distribution of disorderliness can either increase or decrease ductility of the truss lattice metamaterials by a great margin (see SI Appendix, section S1) affecting both stiffness and strength. With the continuation of this, we have developed a physical-based data-driven framework, which tunes the disorderliness to achieve the QTMs with improved ductility. The higher ductility was achieved through changing the failure mechanisms from single shear band formulation to shear band branching or excessive shear tortuosity, which led to desired progressive failure modes. With this data-driven methodology, we can achieve the designs with ductility increased by 30-70% without losing much of their stiffness (<3%) and strength (<13%). Our numerical study has benefited from well designed ANN deep-learning models, built upon a custom-built loss function (SI Appendix, section S3 Fig. S7), which can be trained with a relatively small dataset.

The design of damage tolerant mechanical metamaterials [22] has significant importance in engineering applications. However, prior to this work, there were no deterministic approaches being developed due to indefinite solutions available. We believe that this is just a beginning of an exciting field in the novel topological designs of mechanical metamaterials with tailored properties. Albeit the example we have shown in the result section is based on FCC lattices, the approach is general and applicable to other truss lattice topologies at any scale. The approach proposed in this paper can undoubtedly serve as a unique tool for designing novel mechanical metamaterials well beyond elastic limits.

**Methods**

Details of the creation of FCC-QTM designs and the effects on the level of disorderliness (SI Appendix, section S1); the FE modelling and constitutive damage model (SI Appendix, section S2); the ANN framework, training protocols, accuracy and hyperparameter tuning (SI Appendix, section 3); the effects on the relative density (SI Appendix, section S4); the size effects on macroscopic stiffness and macroscopic peak strength of QTMs (SI Appendix, section S5); the optimization results based on strut thickness variation using strain energy density function (SI Appendix, section S6); and material details used for experimental validation (SI Appendix, section S7) are provided in SI Appendix.

# Supplementary Information Appendix

*S1. Creation of face centre cubic quasi-disordered lattices*

In this section, we provide details to create FCC QTM designs. QTMs can be created via spatial coordinate perturbation of nodes, modelled by introducing geometrical perturbation to the nodes of a perfect FCC periodic lattice. Let $(x^i, y^i)$ represent the spatial coordinates of the $i$th node within a perfect FCC periodic lattice. The new position of the node $(\Delta x^i, \Delta y^i)$ after perturbation can be written as (Fig. S1A):

$$\begin{aligned} \Delta x^i &= \bar{x}^i - x^i = \alpha r \beta \\ \Delta y^i &= \bar{y}^i - y^i = \alpha r \beta \end{aligned} \quad (S11)$$

where $\beta\,(-1 \leq \beta \leq +1)$ denotes a random variable following a uniform probability distribution, $\alpha$ the degree of irregularity related to spatial coordinate perturbation, and $r$ the minimum distance between two nodes within the parent periodic FCC lattice, which can be calculated as:

$$r = \sqrt{\frac{u^2 + v^2}{4}} \quad (S12)$$

where $u$ and $v$ are the lengths of the unit cell in the $x$ and $y$ directions of the parent periodic FCC lattice, respectively (Fig. S1*B*).

Instead, QTMs can be created via variation of strut thickness $(\bar{t})$, which, for the *j*th strut member, can be described as (Fig. S1*B*):

$$\bar{t}^j = (1 + \gamma \beta) t^j \quad (S13)$$

where $\gamma$ the degree of irregularity related to strut thickness variation.

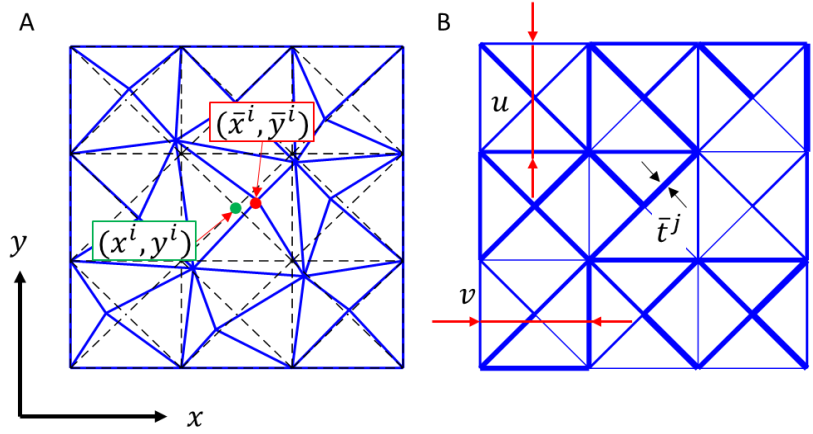

Figure S1 Creation of FCC QTMs via (A) spatial coordinate perturbation (B) strut thickness variation.

*The effects of disorderliness level*

In this section, we posited that it should be possible to design a QTM with high ductility with an increase in disorderliness. The lattice geometry can architect the FCC QTMs by engineering a controlled degree of disorderliness (spatial coordinate perturbations and strut thickness variations) while maintaining a constant relative density (the ratio of density of the lattice material to the density of the solid material from which it is made) $\bar{\rho} = 0.2$. The disorderliness in the FCC QTMs can result in vast differences in ductility, as illustrated in Figs. S2$A$ and $B$. We studied the simulation results for different irregularities, i.e., $\alpha, \gamma = 0.1, 0.2$ and $0.3$. The results are obtained by Finite Element (FE) analysis with 200 calculations for each level of disorderliness, (Figs. S2$A$ and $B$). We identified that FCC QTMs with $\alpha = 0.2$ or $\gamma = 0.1$ have higher ductility advantages without losing much strength ($10-20\%$). From Figs. S2$A$ and $B$, it can be seen that, at a constant disorderliness level, say $\alpha = 0.2$ or $\gamma = 0.1$, the FCC QTMs can fail with either sudden, catastrophic brittle mode or progressive ductile mode.

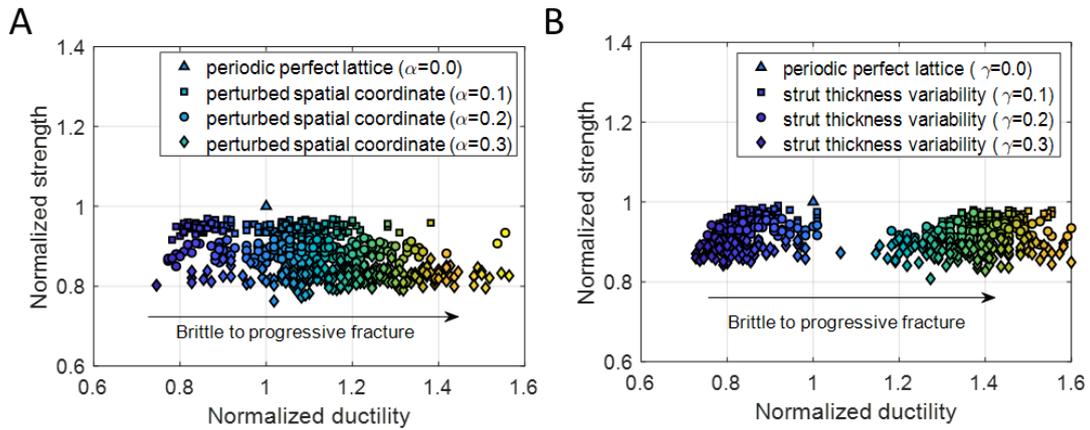

Figure S2 The effects of disorderliness on normalized strength and ductility: (A & B) spatial coordinate perturbation and strut thickness variation, respectively, showing a high range of ductility with increase in disorderliness.

## S2. Finite element modelling and damage model

Here, we present the details for FE modelling on the FCC lattices made of aluminium alloy Al-1050A. The lattice struts were represented as a 2-node Timoshenko-beam element (B21 in ABAQUS notation) with rigid connections. Each strut was modelled numerically as a uniform rectangular cross-sectioned solid bar of in-plane thickness, $t$, and unit out-of-plane width. For the parent periodic FCC lattice with identical lengths of the unit cell in the $x$ and $y$ directions, i.e., $u=v$, the relative density $\bar{\rho}$ of the perfect FCC lattice can be calculated as

$$\bar{\rho} = 2\left(1+\sqrt{2}\right)\left(\frac{t}{v}\right) \tag{S14}$$

The relative density value was kept at $\bar{\rho} = 0.2$ for all QTM topologies in our investigation. Simulation results suggested that converged results could be achieved with each strut meshed with ten beam elements of equal length. To simulate the uniaxial tensile experiment, the specimen was subject to a constant vertical displacement boundary condition on the top and a fixed boundary condition on the bottom, see Fig.S3. The macroscopic stress $\Sigma$, and macroscopic tensile strain $E$ were calculated as:

$$\Sigma = \frac{\text{Reaction Force}}{W},$$
$$E = \frac{\Delta L}{L} \tag{S15}$$

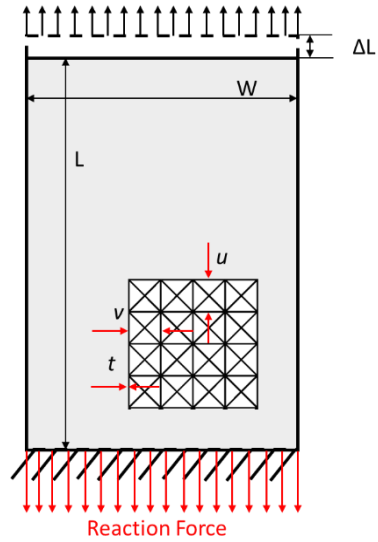

Figure S3 The FE model of a typical metamaterial specimen

The Ramberg-Osgood model was used to represent the true stress-strain relationship of the parent material, i.e., Aluminium alloy Al-1050A, given by:

$$\bar{\varepsilon} = \frac{\bar{\sigma}}{\bar{E}} + \kappa \left(\frac{\bar{\sigma}}{\bar{\sigma}_y}\right)^{\eta} \tag{S16}$$

where $\bar{E} = 70\,\text{GPa}$ and $\bar{\sigma}_y = 134\,\text{MPa}$ are Young's modulus and yield stress of the Aluminium alloy, respectively; $\kappa = 0.002$ is the yield offset and $\eta = 94$ is the hardening exponent [1].

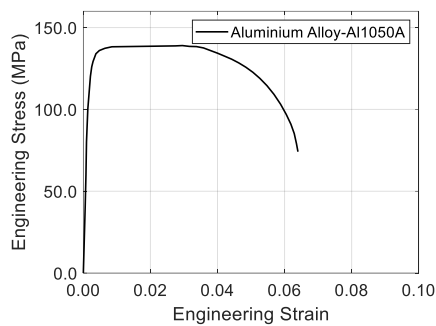

Figure S*4* Engineering stress-strain curve of the aluminium alloy (Al1050A [1]) used for FE simulations

Failure initiation starts when the maximum axial strain reaches 0.03 in the element based on the tensile test result shown in Fig. S4 [1]. The strut necking behaviour is replicated by the reduction of the yield stress after failure initiates, which is characterised by the damage variable *D*:

$$\sigma = (1-D)\sigma_y \tag{S17}$$

where *D* varies from 0 to 1, and is a function of the plastic strain, fitted to match the data of Fig. S4. The corresponding element is deleted from the mesh, when all the material points within the element failed (*D* = 1). Numerical validation was conducted against the experimental data based on a 2D triangular lattice reported by Huaiyuan et al. [1], which suggested that FE simulation could achieve high fidelity.

*S3. Artificial neural network*

In recent studies, various ANN models are now being used. Among the suggested ANN types are multilayer perceptron feed-forward neural networks (FFNN), convolutional neural networks (CNN), and recurrent neural networks (RNN). Each ANN model generally relates to a specific type of issue. FFNN, for example, is widely utilised in many fields and is well-known as "universal approximators" [2]. Compared to CNN and RNN, FFNN has a simpler architecture (only layers and neurons in hidden layers are vulnerable to modification) and is thus easier to evaluate in its diversity. The current study has employed tabular data to relate the input dataset (spatial coordinate perturbations and strut thickness variations) to the relevant output dataset (normalized macroscopic stress-strain response). As a result, an FFNN with a backpropagation algorithm was adopted in this work, as shown in Fig. S5A and B.

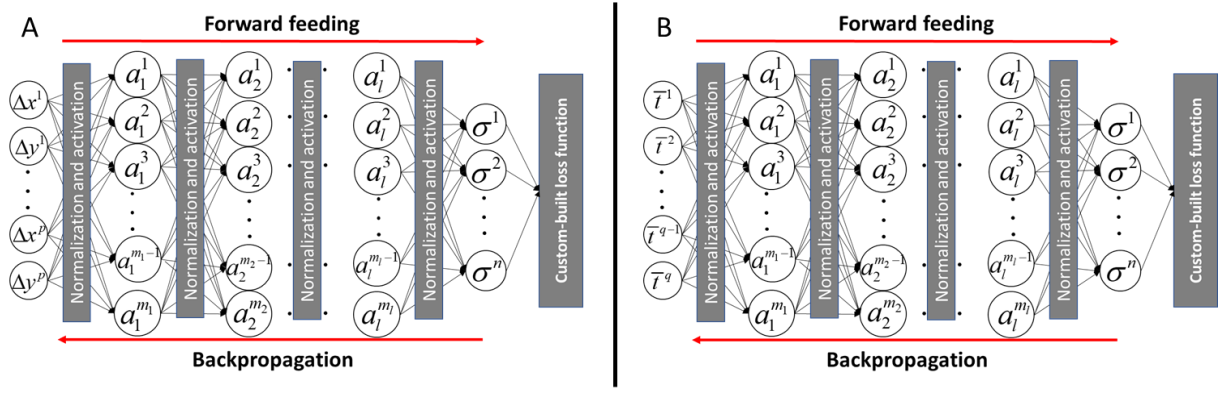

Figure S5 An illustration of feed-forward neural network with backpropogation used in this work (A) spatial coordinate perturbation (B) strut thickness variation

### S3.1. Activation function and normalization

The rectified linear activation function (ReLU) was used in this study. ReLU had been widely used in feed-forward neural networks as an activation function [3]. In this work, the input was normalized between [−1, 1] for spatial coordinate perturbation dataset and [0, 1] for strut thickness variation dataset. It was found that the ReLU activation function performed better with these normalizations. To normalize the $i$th input data, the following mathematical transformation was applied to it:

$$\Psi^{(i)}_{\text{norm.}[-1\ 1]} = 2\frac{\Psi^{(i)} - \min \Psi^{(i)}}{\max \Psi^{(i)} - \min \Psi^{(i)}} - 1, \quad \Psi^{(i)}_{\text{norm.}[0\ 1]} = \frac{\Psi^{(i)} - \min \Psi^{(i)}}{\max \Psi^{(i)} - \min \Psi^{(i)}}, \quad (S18)$$

where $\min \Psi^{(i)}$ is the minimum and $\max \Psi^{(i)}$ is the maximum value of the $i$th component of the input vector $\boldsymbol{\Psi}$ in the dataset.

### S3.2. Evaluation of ANN

The cost function, $J(\boldsymbol{\theta}^{[ii]}, \mathbf{b}_{ii})$, and loss function, $\mathcal{L}(\boldsymbol{\sigma}^m_{pred}, \boldsymbol{\sigma}^m_{true})$, are used to assess the "goodness" of the trained network. The loss function evaluates the model performance based on the real stresses, $\boldsymbol{\sigma}_{true} = \left[\sigma^1, \sigma^2, ..., \sigma^n\right]^{\text{T}}_{true}$, and the predicted stresses, $\boldsymbol{\sigma}_{pred} = \left[\sigma^1, \sigma^2, ..., \sigma^n\right]^{\text{T}}_{pred}$. During training, an optimisation algorithm minimises the value of the loss function by updating the weights and biases values in the "right" direction [4]. The cost function is dependent on the loss function in the following way:

$$J(\boldsymbol{\theta}^{[ii]}, \mathbf{b}_{ii}) = \frac{1}{h}\sum_{m=1}^{h}\mathcal{L}(\boldsymbol{\sigma}^m_{pred}, \boldsymbol{\sigma}^m_{true}) \quad (S19)$$

where $h$ is the number of samples in an evaluated dataset.

The most commonly used loss function is the mean squared error (MSE) for regression analysis problems. The equation is expressed as:

$$\mathcal{L}_{MSE}(\boldsymbol{\sigma}_{pred}^m, \boldsymbol{\sigma}_{true}^m) = \frac{1}{n}\sum_{k=1}^{n}(\sigma_{pred}^k - \sigma_{true}^k)^2 \tag{S20}$$

Apart from this, the logcosh function works similarly to the mean squared error with resistance to outliers, which can be expressed as:

$$\mathcal{L}_{\log\cosh}(\boldsymbol{\sigma}_{pred}^m, \boldsymbol{\sigma}_{true}^m) = \frac{1}{n}\sum_{k=1}^{n}\log(\cosh(\sigma_{pred}^k - \sigma_{true}^k)) \tag{S21}$$

As shown in Fig. S6, numerical experiments on quasi-disordered FCC lattices have suggested that the stress data in the three groups (Zones) have significantly different variances across the QTM samples. Hence, we have proposed a custom-built loss function based on the "quantile regression loss function" to accurately predict stress-strain responses. The loss function used is given as:

$$\mathcal{L}_{custom}(\boldsymbol{\sigma}_{pred}^m, \boldsymbol{\sigma}_{true}^m) = \frac{1}{3n}\sum_{i=1}^{3}\left[\sum_{\substack{k=1 \\ \sigma_{true}^k < \sigma_{pred}^k}}^{n}(\lambda_i - 1)\left(\sigma_{pred}^k - \sigma_{true}^k\right)^2 + \sum_{\substack{k=1 \\ \sigma_{true}^k \geq \sigma_{pred}^k}}^{n}\lambda_i\left(\sigma_{pred}^k - \sigma_{true}^k\right)^2\right] \tag{S22}$$

where $\lambda_i$, $i = 1,...,3$, are the chosen quantiles for the three groups of the stress data and have values between 0 and 1. The quantile loss function is an extension of the Mean Square Error (MSE) that has the quantile $\lambda_i = 0.5$.

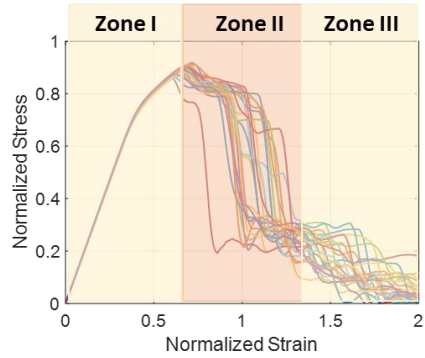

Figure S6 Three zones in the stress-strain relation for the QTM samples under uniaxial tension

### S3.3. ANN Optimisation algorithm

Adaptive Moment Estimation (Adam) [5] is a widely used gradient descent-based backpropagation optimisation algorithm. In our study, this algorithm was used to train ANN models. For each parameter, the algorithm computes adaptive learning rates. It keeps an exponentially decaying average of previously squared gradients $\upsilon_t$, like Nesterov's accelerated gradient method [6], MaxProp [7], and others. However, it differs in the way it updates an exponentially decaying average of past gradients:

$$\zeta_t = \omega_1 \zeta_{t-1} + (1-\omega_1) g_t,$$
$$\upsilon_t = \omega_2 \upsilon_{t-1} + (1-\omega_2) g_t^2,$$
(S23)

where $g_t$ the gradient; $\zeta_t$ and $\upsilon_t$ are approximations of the gradient's first moment (the mean) and second moment (the non-centred variance) at $t$th step. To compensate for moments that are biased towards zero, bias-corrected first and second moment estimates are computed:

$$\hat{\zeta}_t = \frac{\zeta_t}{1-\omega_1^t}$$
$$\hat{\upsilon}_t = \frac{\upsilon_t}{1-\omega_2^t}$$
(S24)

Eventually, parameters are updated according to:

$$\theta_{t+1} = \theta_t - \frac{\chi}{\sqrt{\hat{\upsilon}_t} + \epsilon} \hat{\zeta}_t$$
(S25)

Where $\omega_1$, $\omega_2$, $\chi$, and $\epsilon$ are the algorithm hyperparameters and are subject to tuning.

### S3.4. ANN architecture - hyperparameters

In this section, we have given the hyperparameters used to train our ANN model. Each ANN model received a total of *1000* training iterations. The initial learning rate, $\chi$, is *0.0009*. It decreases by the factor *0.2427* when no training progress is made for 18 consecutive training epochs. The other optimizer hyperparameters were used as their default settings in MATLAB: $\omega_1$ = *0.9*, $\omega_2$ = *0.999*, and $\epsilon = 10^{-8}$. Early stopping was used for deep learning to stop training if the change in learning metrics did not exceed over ten consecutive training iterations. A batch size of 16 was used to train the networks. The hyperparameters to tune the neural networks are obtained using Bayesian optimization from MATLAB (*'bayesopt'*) [8].

### S3.5. ANN architecture analysis

The ANN model was analyzed based on hyperparameters mentioned above. The geometries of 5000 QTM samples were generated with irregularity, $\alpha = 0.2$, at constant relative density $\bar{\rho} = 0.2$. We used an ANN architecture consisting of 7 hidden layers with 4096, 2048, 1024, 1024, 1024, 512, and 512 neurons, in sequence from input to output layers in our architecture. Further increase in hidden layers did not show any improvement in the efficiency of deep learning process. The dataset was split into three sub-datasets 75% for training, 15% for validations and 15% for tests. The evaluation of the three loss functions mentioned in Eqs. S9-11 are presented in the Fig. S7 based on the training and validation datasets. It can be observed that, our custom-built loss function ($\lambda_1$ = 0.5, $\lambda_2$ = 0.45 and $\lambda_3$ = 0.1) has minimum loses compared to the other two loss functions (Eqs. S9 and S10).

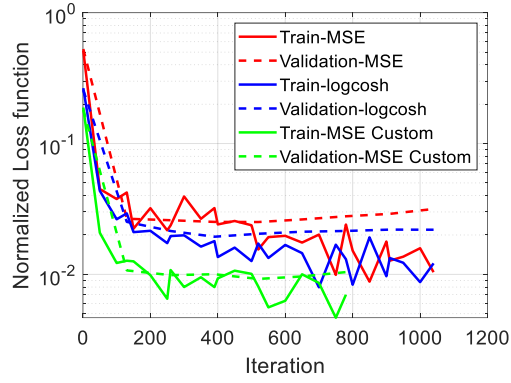

Figure S7 The effects of loss functions on deep learning rate, which shows that the costom built loss function has the best performance

Fig. S8 compares the FEA with the ANN predictions of the stress-strain curves of the FCC QTMs generated via nodal perturbations ($\alpha = 0.2$). In the Fig. S8, the first, second, and third row plots compare training, validation, and test datasets, respectively. The stress-strain curves are randomly selected from the respective datasets. Similarly, Fig. S9 compares the FEA with the ANN predictions for FCC QTMs generated via strut thickness variations ($\gamma = 0.1$).

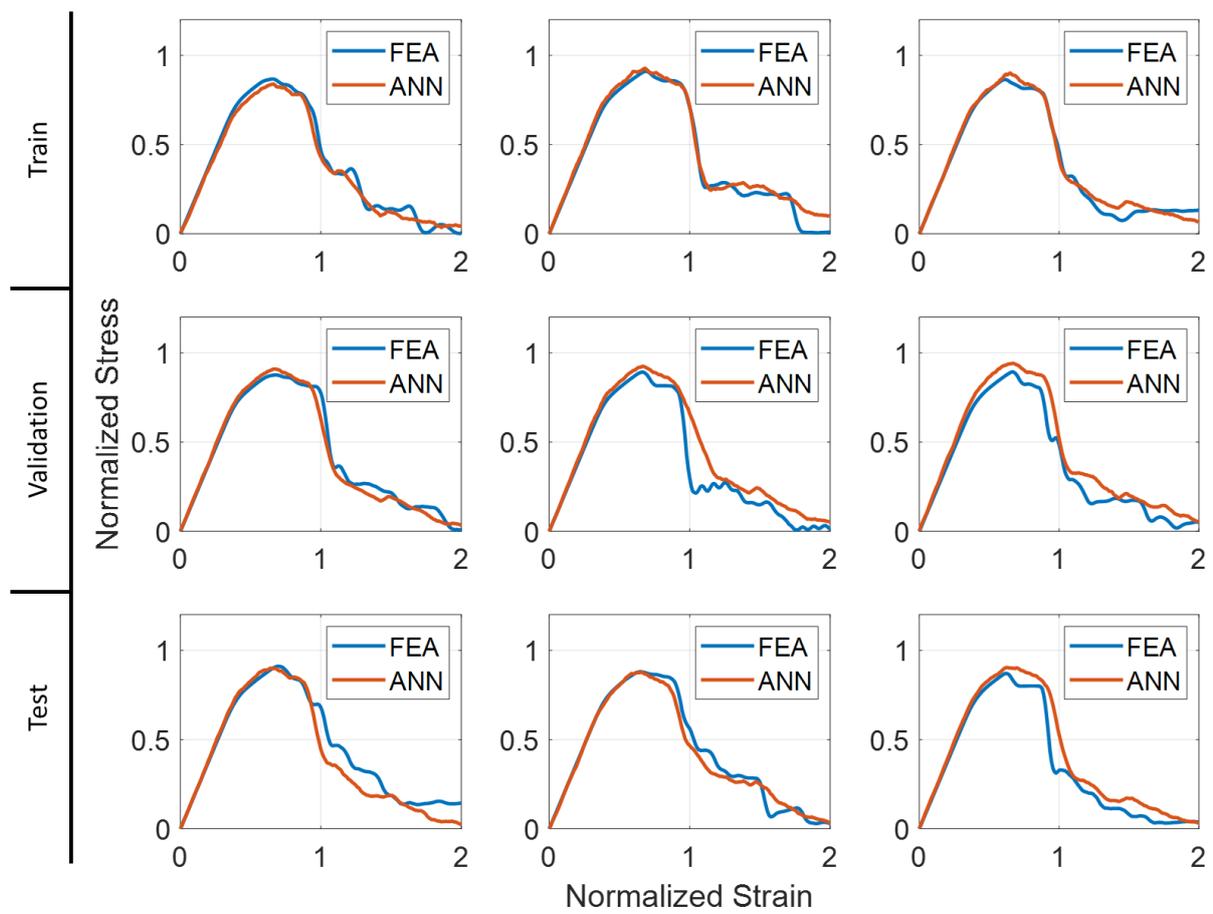

Figure S8 The comparison of the FEA results with the ANN predictions for FCC QTMs generated via spatial coordinate perturbations

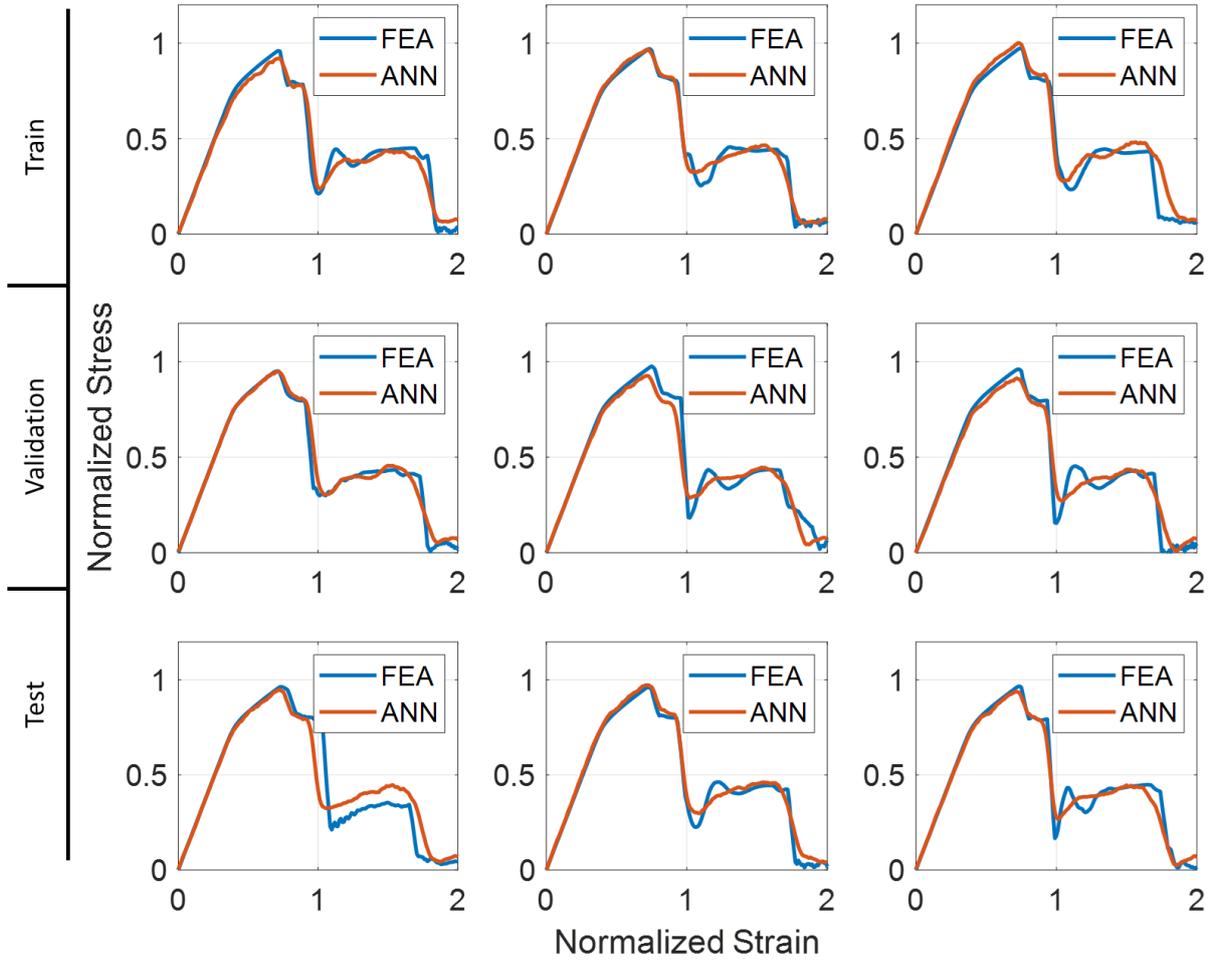

Figure S9 The comparison of the FEA results with the ANN predictions for FCC QTMs generated via strut thickness variations

## S4. Effects of relative density of QTMs

In this section, the investigation of relative density has been carried out. We have conducted FE simulations for 100 QTMs at relative densities of $\bar{\rho} = 0.2, 0.3$ and $0.4$, respectively, with nodal perturbation irregularity $\alpha = 0.2$. The normalized histogram plots for ductility, strength and stiffness variation are given in Fig. S10. Figs. S11*A, B* and *C,* show the comparison of ductility, strength, and stiffness variability for relative density, $\bar{\rho} = 0.2, 0.3$ and $0.4$, respectively. It is evident that $\bar{\rho}$ =0.2 shows the highest variability in ductility while the change in strength remains constant across chosen relative densities, see Fig. S11B. It can also be noted that the change in stiffness is increasing as we increase the value of relative densities, Fig. S11*C*. Note that the effects of relative density may vary depending on the irregularity values.

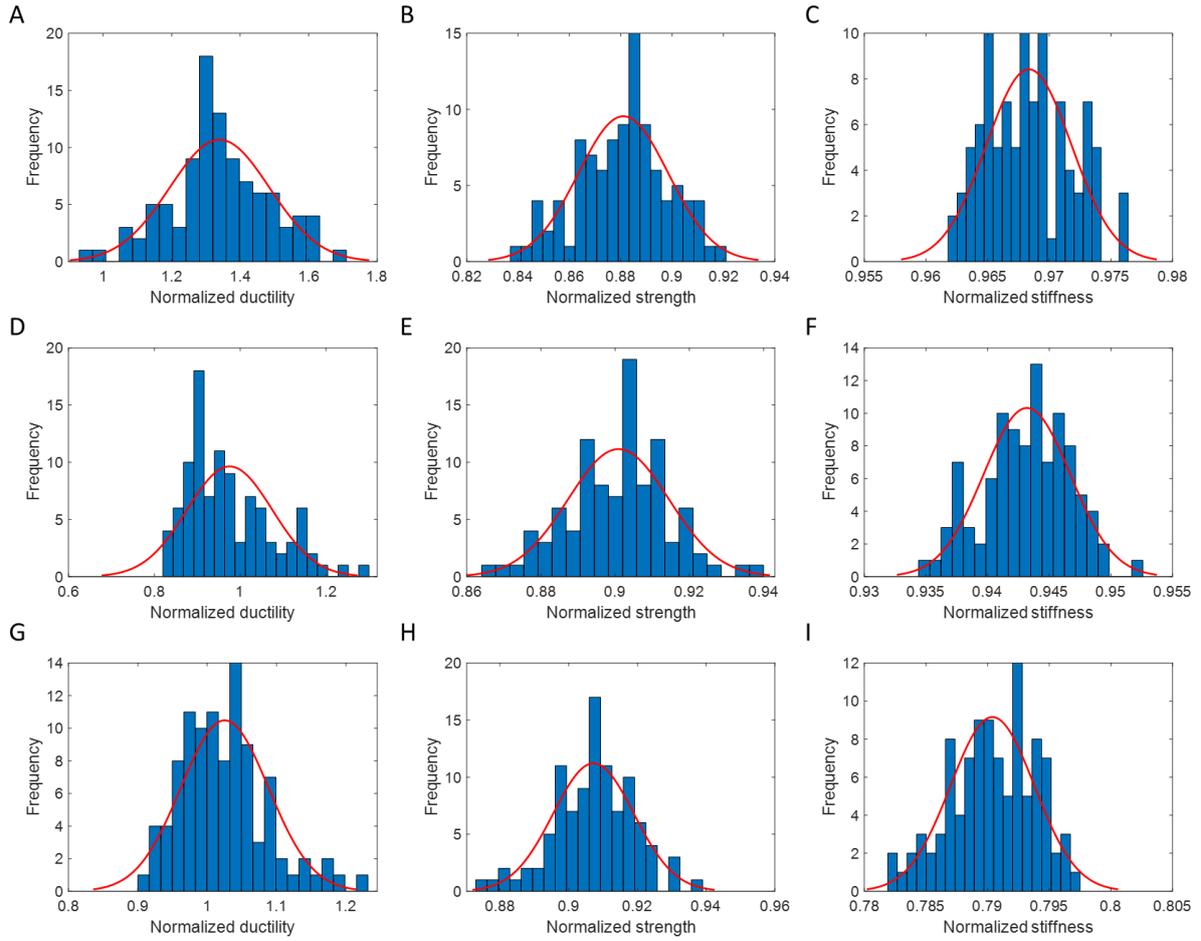

Figure S10 The histogram plots for ductility, strength and stiffness variation normalized with respect to perfect periodic FCC lattice, (A, B and C) for $\bar{\rho}=0.2$, (D, E and F) for $\bar{\rho}=0.3$ and (G, H and I) for $\bar{\rho}=0.4$, respectively.

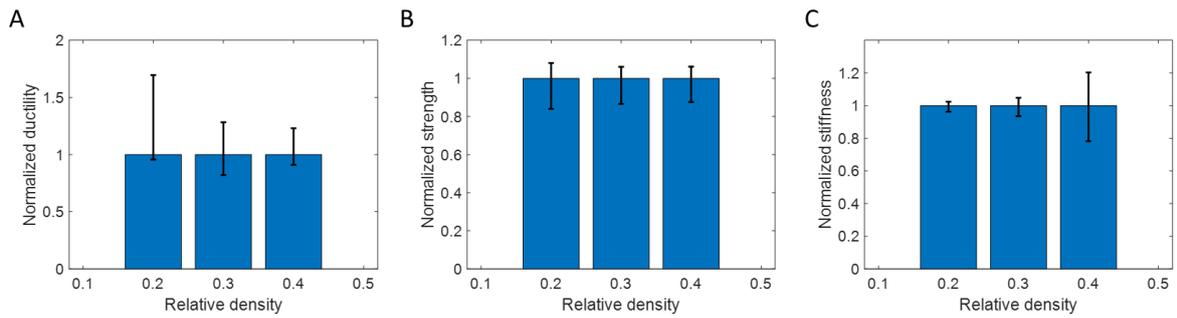

Figure S11 The comparison of (A) ductility, (B) strength, and (C) stiffness variability for relative density, $\bar{\rho}$, 0.2, 0.3 and 0.4, respectively. The values are normalized with respect to perfect periodic FCC lattice.

*S5. Size effects of FCC QTMs*

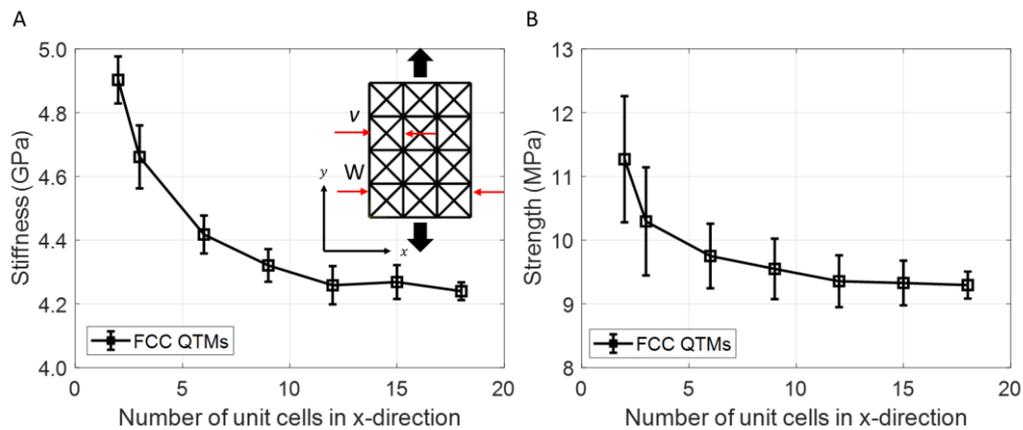

Figure S12 The size effects on (A) the structural macroscopic stiffness and (B) the structural peak strength of the FCC QTMs generated via spatial coordinate perturbations

The size effects on the macroscopic stiffness and the macroscopic peak strength of the FCC QTMs were investigated using FE simulations. The details of the FE simulations are described in SI Appendix S2. The QTMs were created based on the perfect FCC lattice with square unit cells (i.e., $u=v$). The width to height ratio, W/L, of the QTM samples were kept at 0.75; and the unit cell size, $v$, and the thickness, $t$, were taken as 10 mm and 0.4 mm, respectively. The size effects were evaluated by increasing the number of unit cells from 2 to 18 in $x$-direction. We have conducted FE simulations for 100 QTMs at relative densities of $\bar{\rho}=0.2$ with nodal perturbation irregularity $\alpha=0.2$ for each sample size. Fig. S12A shows the size effect via the functional relationship of the structural macroscopic stiffness against the number of unit cells in $x$-direction. The macroscopic stiffness is not sensitive to the number of unit cells when number of cells were more than 12. As shown in Fig. S12B, the peak strength converged at the lattice size of 12 unit cells in $x$-direction (i.e., 16 unit cells in $y$-direction). Thus, we opted for the QTMs of $12\times16$ unit cells for this methodology development, provided in the main text.

*S6. The optimized results – strut variation based on the strain energy density function*

In this case, the disorderliness was introduced using strut thickness variation to the perfect periodic FCC lattices. Based on the ANN model and three selected initial distribution of disorderliness, we have obtained three optimized QTM designs using the strain energy density objective function, as shown in Fig. S13. We have obtained the optimum designs that exhibited progressive failure modes compared to brittle failure of the perfect periodic FCC lattice (Fig. 3A, in the main text). The progressive failure modes in the optimized designs were mainly achieved by the shear band branching, as shown in Fig. S13D. The optimized designs exhibit more than 60% increase in strain energy density with the expense of less than 10% strength and less than 5% stiffness (Figs. S13 A, B and C). It is apparent that all optimized QTMs show distinct failure mechanisms, while achieving higher strain energy density.

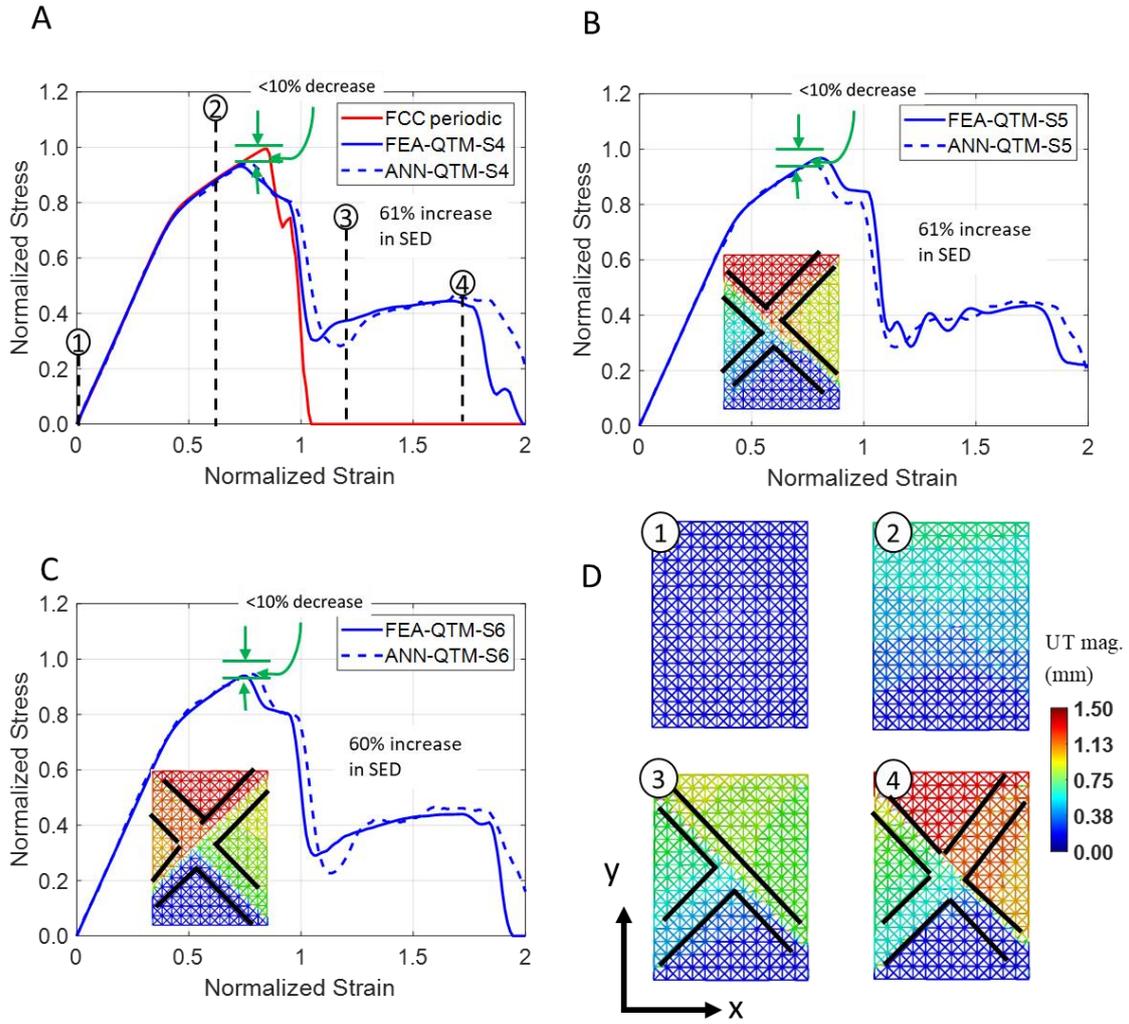

Figure S13 Designs of metamaterials based on strut thickness variation using the strain energy density objective function, (A, B and C) the normalized stress-strain curves of three optimized QTMs (-S4, -S5 and -S6) obtained by the FE simulations and ANN predictions; (D) the distribution of displacements at selected macroscopic strains of a QTM (-S4), showing that shear band branching causes progressive failure mode. The inserts of (B, and C) show the distribution of the displacement at failure, which caused by the formulation of the shear band branching with different patterns.

## S7. Stress-strain curves of polymers

Fig. S14A shows the engineering stress-strain curves of *polymer1* and *polymer2* used in the experimental study. The optimized QTM design for the Experimental Study in the main text is manufactured using these two materials. A miniature specimen tensile tests were performed to get engineering stress-strain curves. The geometrical dimensions of the miniature sample are shown in Fig. S14B. The detailed geometry of the experimental sample given in Fig. S15 (Fig. 3H in the main text) which was obtained after optimization and used for the tensile test.

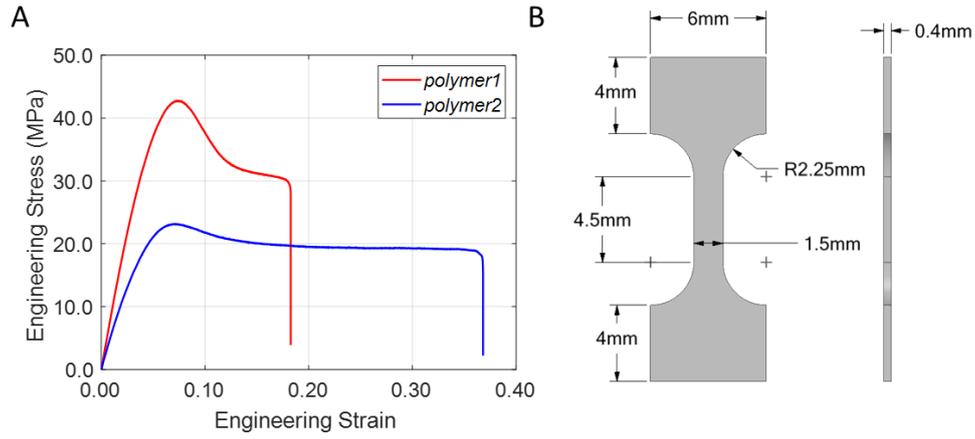

Figure S14 (A) engineering stress-strain curves of the *polymer1* amd *polymer2*; (B) miniature specimen dimensions

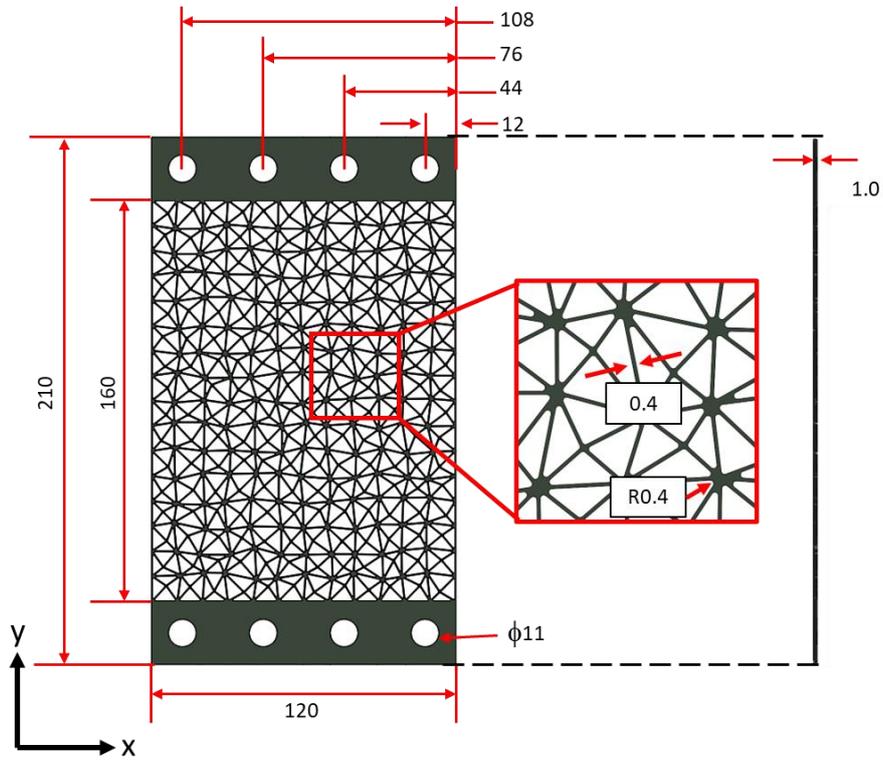

Figure S15 Geometry details of the optimized QTM for uniaxial tensile test (all dimensions are in mm)